\documentclass[12pt,twoside]{article}   
\usepackage{comment}
\usepackage{caption}
\usepackage{subcaption}
\usepackage{amsmath}
\usepackage[super,sort&compress,comma]{natbib}

\usepackage{fancyhdr}		

\usepackage[section]{placeins}   %

\usepackage{graphicx}

\makeatletter \renewcommand\@biblabel[1]{$^{#1}$} \makeatother
\newcommand{\cen}[1]{\begin{center} #1 \end{center}}

\usepackage[mathlines]{lineno}

\usepackage{hyperref}
\hypersetup{ colorlinks,
    citecolor=blue,
    filecolor=blue,
    linkcolor=blue,
    urlcolor=blue
}

\usepackage{xcolor}

\definecolor{gray}{rgb}{0.6,0.6,0.6}
\definecolor{red}{rgb}{0.85,0,0}
\definecolor{green}{rgb}{0,0.85,0}
\definecolor{blue}{rgb}{0,0,0.85}
\definecolor{beige}{rgb}{0.92,0.87,0.78}
\usepackage[all]{hypcap}    

\begin{document}

\cen{\sf {\Large {\bfseries Human-like AI-based Auto-Field-in-Field Whole-Brain Radiotherapy Treatment Planning With Conversation Large Language Model Feedback} \\  
\vspace*{10mm}
Adnan Jafar$^1$, An Qin$^1$, Gavin Atkins$^1$, Xiaoyu Hu$^1$, Yin Gao$^2$, Xun Jia$^{1*}$} \\
1. Department of Radiation Oncology and Molecular Radiation Sciences, Johns Hopkins University, Baltimore, MD, USA\\
2. Department of Radiation Oncology, University of Pennsylvania, Philadelphia, PA, USA\\
\vspace{5mm}
Version typeset \today
}

\pagenumbering{roman}
\setcounter{page}{1}
\pagestyle{plain}
* Author to whom correspondence should be addressed. Email: xunji@jhu.edu \\

\begin{abstract}

{\bf Background:} Whole-brain radiotherapy (WBRT) is a common treatment due to its simplicity and effectiveness. While automated Field-in-Field (Auto-FiF) functions assist WBRT planning in modern treatment planning systems, it still requires manual approaches for optimal plan generation including patient-specific hyperparameters definition and plan refinement based on quality feedback.\

{\bf Purpose:} This study introduces an automated WBRT planning pipeline that integrates a deep learning (DL) Hyperparameter Prediction model for patient-specific parameter generation and a large-language model (LLM)-based conversational interface for interactive plan refinement.\

{\bf Methods:} The Hyperparameter Prediction module was trained on 55 WBRT cases using geometric features of clinical target volume (CTV)  and organs at risk (OARs) to determine optimal Auto-FiF settings in RayStation treatment planning system. Plans were generated under predicted hyperparameters. For cases in which the generated plan was suboptimal, quality feedback via voice input was captured by a Conversation module, transcribed using Whisper, and interpreted by GPT-4o to adjust planning settings. Plan quality was evaluated in 15 independent cases using clinical metrics and expert review, and model explainability was supported through analysis of feature importance.\

{\bf Results:} Fourteen of 15 DL-generated plans were clinically acceptable. Normalized to identical CTV D95\% as the clinical plans, the DL-generated and clinical plans showed no statistically significant differences in doses to the eyes, lenses, or CTV dose metrics D1\% and D99\%. The DL-based planning required under 1 minute of computation and achieved total workflow execution in approximately 7 minutes with a single mouse click, compared to 15 minutes for manual planning. In cases requiring adjustment, the Conversational module successfully improved dose conformity and hotspot reduction.\

{\bf Conclusions:} The proposed system improves planning efficiency while maintaining clinically acceptable plan quality. It demonstrates the feasibility of combining DL-based hyperparameter prediction with LLM interaction for streamlined, high-quality WBRT planning.\

\end{abstract}

\newpage
\tableofcontents

\newpage

\setlength{\baselineskip}{0.7cm}      

\pagenumbering{arabic}
\setcounter{page}{1}
\pagestyle{fancy}

\section{Introduction}

Brain metastases occur in approximately 25\% of cancer patients, with lung and breast cancers being the most common primary origins \cite{Wen2001}. There has been an increase in brain metastases driven by improved cancer treatments prolonging survival, advancements in imaging enabling earlier detection, and the rising prevalence of highly metastatic cancers like melanoma \cite{McTyre2013}. Brain metastases are typically managed palliatively \cite{Sharma2008, Nieder2010}, with treatment options including chemotherapy, surgery, radiotherapy, or a combination of these approaches. Stereotactic radiosurgery (SRS) and whole-brain radiotherapy (WBRT) are common radiotherapy options for managing brain metastases. For patients with a single operable brain metastasis, surgery followed by WBRT is a standard treatment that has been shown to improve survival and oncological outcomes \cite{Yu2021, Lamba2017, Brown2016}. SRS offers better local control and fewer side effects than WBRT for up to three metastases but has a higher risk of recurrence outside the treated area \cite{Kocher2011, Aoyama2006, Yamamoto2014}. Despite various treatment approaches\cite{Andrews2004}, WBRT remains a widely used treatment option for brain metastases \cite{Andrews2004, Franchino2018}.

WBRT is often delivered using two parallel opposed lateral beams with the patient in the supine position \cite{McTyre2013, Sharma2008}. The goal of WBRT treatment planning is to generate a treatment plan with a homogeneous dose coverage to the Clinical Target Volume (CTV), typically including whole-brain parenchyma, while minimizing radiation doses to critical organs, such as lenses, optic nerves/chiasm, etc. In the current practice, human planners develop the WBRT treatment plan typically using a forward planning method in collaboration with physician feedback. A field-in-field (FiF) technique is often employed because of its simplicity, robustness to setup variation, and delivery with the standard step-and-shoot approach \cite{Kim2018, Huang2023}. The planning process of FiF for WBRT involves defining contours for organs and the CTV, setting the isocenter, and specifying the prescription dose. Human planners begin by defining the aperture formed by the multi-leaf collimator (MLC) for two opposing lateral beams, and then iteratively add segments to each beam to reduce hot spots, manually adjusting the monitor units (MUs) for each segment until the desired plan is achieved. To streamline this process, commercial treatment planning systems (TPSs) have implemented Auto-FiF algorithms that iteratively add beam segments for two opposing lateral beams, aiming to block hot spots and minimize the low-dose regions through beam weight optimization \cite{Byrd1995}. The Auto-FiF algorithm has been successfully evaluated in various disease sites, including whole-brain, rectal, cervical, and breast \cite{Huang2023, Huang2022, Huang2021, Onal2012}. 

Despite the success of the Auto-FiF algorithm in automating the planning process, two major factors affect its efficacy. The first is the choice of hyperparameters in the Auto-FiF algorithms. These hyperparameters define the user’s preferences and guidance on various aspects of the resulting plan, thereby critically influencing its quality and clinical acceptability. Taking the RayStation TPS (RaySearch Laboratory, Stockholm, Sweden) as an example, these parameters include target coverage priority, number of subfields, minimum segment monitor unit (MU) per fraction, minimum segment area, minimum number of open leaf pairs, and minimum leaf end separation. Improper settings can lead to sub-optimal plan quality. The second factor arises after the initial plan is generated with the Auto-FiF algorithm. If further refinements are needed based on the physician’s input, it is critically important to streamline the process of updating the plan so that the TPS can revise it in the desired direction.

Recent advancements in deep learning (DL) have significantly transformed RT treatment planning \cite{li2020automatic, Gao2024Human, jia2025nrg, shen2020introduction, shan2020synergizing}. DL-based methods have been proposed to automate the WBRT planning \cite{Yu2021, Han2021}. These methods are based on either predicting the beam apertures on lateral-opposed fields \cite{Yu2021} or the MLC positions for two lateral beams \cite{Han2021}. In this study, we will explore the use of DL to facilitate the Auto-FiF function by generating patient-specific planning hyperparameter settings of the Auto-FiF algorithm in the RayStation TPS. Similar to the research direction on building a virtual planner to operate a TPS \citep{Gao2024Human,shen2021hierarchical}, this approach allows us to employ a commercial TPS to directly generate plans, while letting the TPS manage the detailed but critical physics modeling part of the plan, such as dose calculation and MLC modeling, ensuring plan deliverability. 

To streamline the process of plan refinement, we propose the use of modern AI tools for human-AI interaction. Recently, large language models (LLMs) have shown significant potential in RT treatment planning \cite{liu2024automated, Wang2025}, owing to their zero-shot capabilities learned through self-supervised training on large-scale, diverse corpora, followed by alignment with human preferences \cite{Naveed2023}. Their ability to understand, interpret, and summarize unstructured clinical input into structured representations makes them valuable for various decision-support tasks in healthcare \cite{Naveed2023, Thirunavukarasu2023}. Accordingly, in this study, we will develop a human-AI Conversation module to instruct the planning tool, enabling the seamless intake of human instructions and facilitating the generation of plans that meet clinical requirements.

\section{Methods}

\begin{figure}[tbp]
    \centering
    \includegraphics[width=0.8\linewidth]{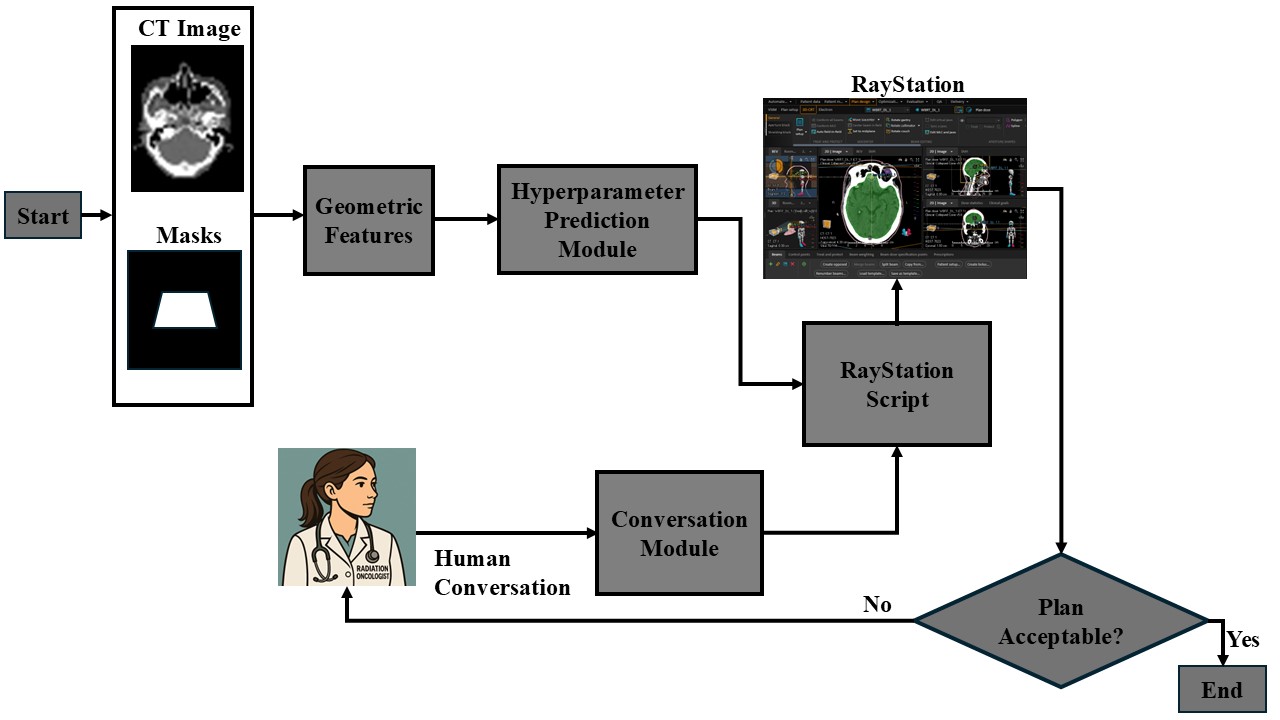}
    \caption{Workflow of the Auto-FiF treatment planning for WBRT, augmented with an explainable DL-based Hyperparameter Prediction module and a Conversation module powered by an LLM.}
    \label{fig:workflow}
\end{figure}

The Auto-FiF algorithm implemented in commercial TPSs relies on a set of hyperparameters that directly govern plan quality. In routine practice, specifying these hyperparameters is a manual process that is iterative and sometimes necessitates repeated adjustments. This reliance on manual fine-tuning limits efficiency and introduces variability across planners. To address these challenges, we designed a fully automated workflow for WBRT planning based on Auto-FiF, with the workflow illustrated in Figure~\ref{fig:workflow} that consists of two tightly integrated modules: Hyperparameter Prediction module and Conversation module.

The Hyperparameter Prediction module employs a supervised DL approach to predict optimal Auto-FiF hyperparameter settings in the TPS. The inputs are geometric descriptors of CTV and adjacent organs at risk (OARs), which are mapped to hyperparameter values that can emulate physician-approved clinical WBRT plans. We selected a supervised learning framework, because the task is explicitly defined, making it well-suited to supervised regression/classification methods. Geometric features, rather than full CT volumes, were deliberately chosen as inputs, because WBRT involves relatively simple anatomical geometry, where spatial relationships between target and organs are expected to provide sufficient guidance for planning. This choice also enhances model explainability and reduces computational burden. To further strengthen transparency and clinical trust, the module incorporates an explainability layer that highlights which geometric features most strongly influence the predicted hyperparameters. Based on the predicted hyperparameters, the Auto-FiF function was launched to generate a plan. 

When this automatically generated plan requires further improvement to meet clinical expectations, the Conversation module provides a novel physician–TPS interaction channel. Powered by an LLM, this module allows physicians to directly articulate requested modifications in natural language, e.g., “reduce hotspots in the parietal region” or “improve homogeneity across the left temporal lobe”. The system translates these instructions into actionable adjustments within the TPS. This direct pathway not only accelerates iteration but also reduces communication bottlenecks and enhances physician oversight of final plan quality.

The subsequent sections present the technical implementation of these two modules, highlighting their design principles, integration within the TPS, and evaluation in clinical WBRT scenarios.

\subsection{Hyperparameter Prediction module}
\subsubsection{Dataset}

We retrospectively selected 70 patients who had previously undergone WBRT at our clinic. For each case, treatment plan data in DICOM format were exported from RayStation TPS, including CT images, RTstruct, and RTplan files. Because the Hyperparameter Prediction module was formulated as a supervised learning task, the dataset was constructed to include geometric features as the module's inputs and the corresponding ground-truth Auto-FiF hyperparameter settings as outputs. To obtain these ground-truth hyperparameters for output, we employed a Python script that systematically executed the Auto-FiF algorithm in RayStation TPS across a range of candidate hyperparameter settings, using a grid-search strategy to identify the hyperparameter configuration that best reproduced the physician-approved clinical plan. Through this process, we identified four key parameters: target coverage priority (25 to 100\%, step size 25\%), number of subfields (1 to 3, step size 1), minimum segment MU per fraction (2 to 8~MU, step size 2), and minimum segment area (4, 6, 9, and 10~cm$^{2}$), which required adjustment to faithfully replicate the clinical plans. The remaining parameters, such as the minimum number of open leaf pairs \((1)\) and minimum leaf end separation \((0\,\mathrm{cm})\), were consistently kept at their default values for all cases.

\subsubsection{Network Structure}

\begin{figure}[htbp]
    \centering
    \includegraphics[width=0.8\linewidth]{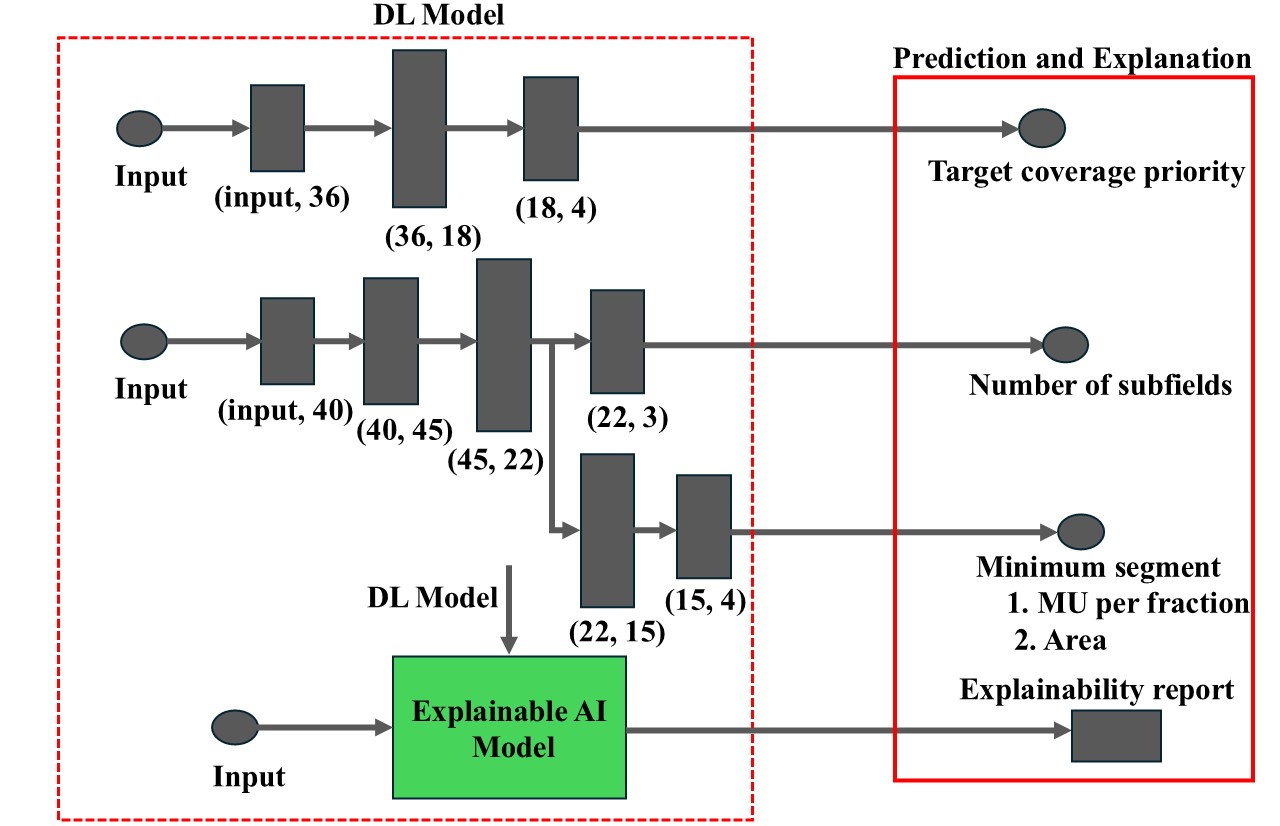}
    \caption{Architectures of the DL models for Auto-FiF hyperparameter prediction, including a model for target coverage priority, a model for the number of subfields, and the third one that reused the subfield classifier backbone with additional layers to predict minimum segment MU per fraction and minimum segment area. An explainable AI module was employed for attributing model predictions to geometric features.}
    \label{fig:network_a}
\end{figure}

We formulated the task of learning the optimal Auto-FiF hyperparameters as a multi-class classification task driven by geometric features of the target and surrounding anatomy, as shown in Figure~\ref{fig:network_a}. To this end, we designed three DL models with fully connected architectures. The first model predicted the target coverage priority, discretized into classes ranging from 25\% to 100\% in 25\% increments. The second model predicted the number of subfields (1, 2, or 3). The third model, which shared its backbone architecture with the number of subfield-prediction network, predicted a joint label encoding combinations of the minimum segment MU per fraction (2, 4, 6, or 8) and the minimum segment area (4, 6, 9, or 10). For the third model, we limited the label space to four clinically relevant combinations: (2, 4), (4, 6), (6, 9), and (8, 10), as these parameter sets were consistently sufficient to reproduce physician-approved clinical plans.

\subsubsection{Geometric Features}

\begin{figure}[htbp]
    \centering
    \includegraphics[width=0.9\linewidth]{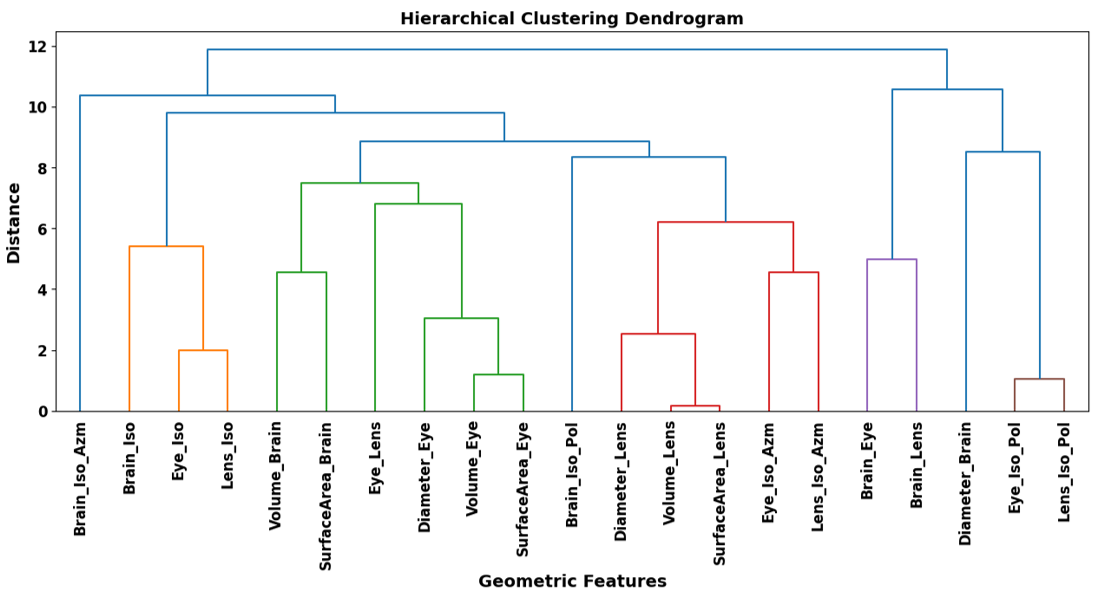}
    \caption{Dendrogram illustrating the hierarchical clustering of geometric features to identify feature relationships, guiding the iterative feature selection process for optimizing model performance.}
    \label{fig:dendogram}
\end{figure}

We selected the brain, left and right eyes, and left and right lenses as the organs of interest, consistent with the structures typically evaluated by physicians when generating treatment plans for WBRT. Instead of inputting the 3D contours of these structures directly to the network for predictions, we extracted relevant geometric features out of these structures to avoid overfitting under the relevant small data size and support explainability. The geometric features initially chosen included the volume, surface area, diameter of each organ, centroid-to-centroid inter-organ distances, distance from the isocenter to each organ's centroid, as well as the polar and azimuthal angles from the isocenter to each organ's centroid. For features with lateral symmetry, e.g., distances between the isocenter to left and right eyes, the corresponding feature values were averaged to remove redundancy among the candidate features. This resulted in a total of 21 features, as shown in the bottom of Figure~\ref{fig:dendogram}. 

To remove redundant information and focus on most relevant information offered by the set of the geometric features, we performed agglomerative hierarchical clustering \cite{Mullner2011} on the standardized input features (Figure~\ref{fig:dendogram}), employing the Euclidean distance metric. Agglomerative hierarchical clustering iteratively merged clusters based on chosen linkage criteria and distance metrics, enabling flexible feature grouping \cite{Mullner2011}. By recursively cutting the dendrogram at specific horizontal thresholds, we iteratively selected features that enhance model performance. Features that did not contribute to improving the classifier's accuracy were discarded, and this process continued until the optimal feature subset was identified, yielding improved model performance. Finally, the geometric features selected for DL model training included the diameters of the brain, eyes, and lenses (\textit{Diameter\_Brain}, \textit{Diameter\_Eye}, \textit{Diameter\_Lens}); the distances from the isocenter to the centroids of the brain, eyes, and lenses (\textit{Brain\_Iso}, \textit{Eye\_Iso}, \textit{Lens\_Iso}); and the polar and azimuthal angles from the isocenter to the centroids of the eyes and lenses (\textit{Eye\_Iso\_Pol}, \textit{Lens\_Iso\_Pol}, \textit{Eye\_Iso\_Azm}, \textit{Lens\_Iso\_Azm}). This result suggests that the sizes of these organs, along with their relative positions to the isocenter, provide unique information that is important for treatment planning.



\subsubsection{Network Training}

We trained three DL models on 55 WBRT cases for 1500 epochs with a batch size of 5, optimizing cross-entropy loss. Five-fold cross-validation was performed on the training cohort to support model selection and mitigate overfitting. Early stopping was monitored through validation loss. All models were trained using the Adam optimizer, which was selected for its adaptive learning rate and robustness to sparse gradients. An independent set of 15 WBRT cases, withheld from both training and validation, was used to evaluate final model performance.

For the target coverage priority model, we applied a dropout rate of 35\% after the first fully connected layer to suppress overfitting, given the relatively small dataset size. The hidden layers employed a combination of Leaky ReLU and ReLU activation functions, chosen to balance gradient stability in early layers with computational efficiency in later layers. The learning rate and weight decay were both set to 0.0002, values that provided stable convergence during pilot experiments.

For the number of subfields model and the minimum segment MU–segment area model, a more conservative regularization scheme was applied, with dropout rates of 10\% after the first layer and 15\% after the second layer. These models also used Leaky ReLU activation in the first layer, followed by ReLU activations in subsequent layers, to promote nonlinearity while maintaining stability. The learning rate and weight decay were set to 0.001 and 0.00009, respectively, to allow faster convergence while preventing overfitting to training data.
 
\subsection{Conversation module}

\begin{figure}[tbp]
    \centering
    \includegraphics[width=0.8\linewidth]{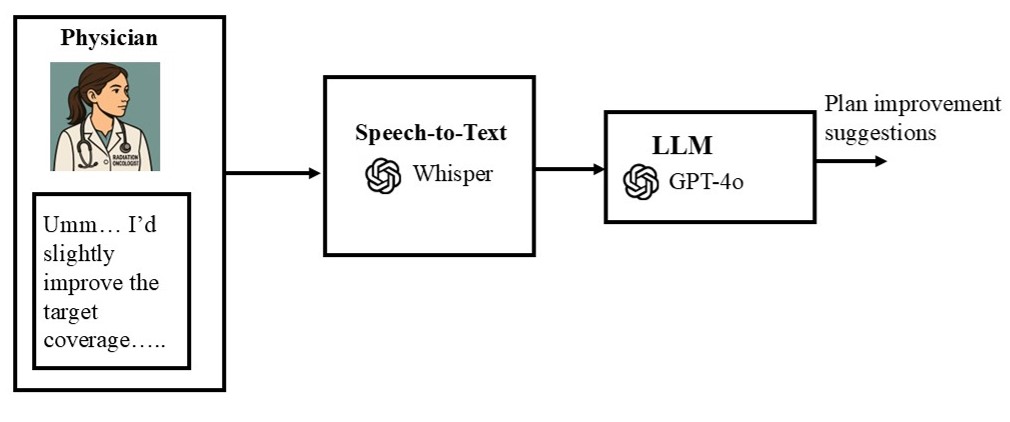}
    \caption{Conversation module for physician-in-the-loop planning. Verbal feedback from the physician was converted to text using Whisper, then interpreted by GPT-4o to produce plan improvement suggestions that align with the physician’s intent.}
    \label{fig:network_b}
\end{figure}

For the Conversation module, physician input was captured, transcribed, interpreted, and translated into optimization objectives in the TPS. Physician speech was first recorded automatically using the \textit{sounddevice} library (a Python package for 
audio recording) and then transcribed into text using OpenAI Whisper, a state-of-the-art speech-to-text model known for its robustness to accents and clinical terminology. The transcribed text was subsequently processed by the GPT-4o API (temperature = 0.2), which generated structured JSON output according to a predefined schema (Figure~\ref{fig:network_b}). The low temperature setting was chosen to minimize randomness and ensure consistent, reproducible outputs suitable for clinical decision-making.

A key design choice was the use of few-shot prompting with GPT-4o. In this strategy, the model was provided with several examples of input–output pairs that demonstrated the desired mapping between physician instructions and structured JSON. For instance, the feedback “The target coverage is poor” was mapped to \texttt{\{ "target\_coverage": "bad", "OAR\_dose": "unchanged" \}}. These exemplars guided the model to consistently interpret clinical feedback in a structured and machine-readable format. This approach was critical because physician instructions are often heterogeneous in phrasing, yet the TPS requires standardized commands to adjust hyperparameters. The feedback could include requests such as improving target coverage, reducing dose in a specific OAR, or modifying the weight of the initial control point beams.

Once generated, the structured JSON was parsed and directly translated into optimization objectives. For example, a request to improve target coverage was operationalized as increasing D\textsubscript{95} or D\textsubscript{99} by at least 5\% compared with the initial plan. To achieve these objectives, the system performed a localized grid search around the hyperparameter values predicted by the Hyperparameter Prediction module. For computational efficiency, the localized grid search was restricted to two parameters, target coverage priority and the number of subfields, since these parameters directly control target coverage and hotspot reduction. The search was initiated with target coverage priority, and if this alone was insufficient, it was extended to include both parameters to achieve physician-preferred trade-offs. If no satisfactory configuration was found, the plan was reverted to that obtained from the Hyperparameter Prediction module.

This design effectively closed the feedback loop by allowing physicians to provide natural, conversational guidance that was reliably converted into actionable optimization strategies, bypassing the traditional, iterative physician–dosimetrist communication cycle.

\subsection{Evaluations}

Evaluation on the Hyperparameter Prediction module was performed at three complementary levels to comprehensively assess both the technical accuracy of the DL models and the clinical utility of the overall workflow. First, we assessed the predictive performance of each multiclass DL module on the independent test set of 15 WBRT cases. The evaluation metrics included precision, recall, and F1-score, which were computed for each class and then macro-averaged across all classes to avoid bias toward more frequent classes. This analysis quantified how reliably the models could predict the discrete hyperparameter settings required for Auto-FiF planning. Second, we conducted a quantitative evaluation of the resulting treatment plans using standard dosimetric metrics, after normalizing the DL-generated plans to the same D95\% as the clinical plans for each case. For the CTV (brain), D1\% and D99\% were evaluated, while for the OARs (eyes and lenses), the mean dose was evaluated. By comparing these dosimetric indices between the DL-augmented Auto-FiF plans and the physician-approved clinical reference plans, we were able to assess how closely the automated workflow reproduced clinically acceptable dose distributions and whether it maintained the delicate balance between target coverage and normal tissue sparing. Finally, a qualitative expert review was performed by experienced medical physicists, who examined the DL-augmented Auto-FiF plans for overall clinical acceptability on a 5-point scale (5 = clinically acceptable, 4 = acceptable with minor edits after normalization, 3 = better than starting from scratch, 2 = major edits required, 1 = unusable). This review focused not only on objective criteria such as coverage and OAR sparing but also on subjective aspects such as the presence of hotspots in sensitive anatomical regions, conformity of dose distribution to clinical expectations, and general plan quality. 

The Conversation module was evaluated through a combination of functionality testing, plan adjustment accuracy, and expert review. To demonstrate its ability to interpret physician feedback, we designed a set of representative clinical scenarios in which users provided spoken instructions, under which the plans were refined. The resulting Auto-FiF plans were then compared against baseline plans generated by the predicted hyperparameter, to determine whether the requested modifications were successfully incorporated. Quantitative evaluation focused on whether targeted metrics improved in line with the users' request, while qualitative assessment was conducted by a medical physicist to confirm that the dose distribution changes were clinically meaningful and did not compromise other aspects of plan quality. 

\section{Results}

\subsection{Hyperparameter Prediction module}

\subsubsection{DL-augmented Auto-FiF Plans} 

\begin{figure}[htbp]
    \centering
    \includegraphics[width=\linewidth]{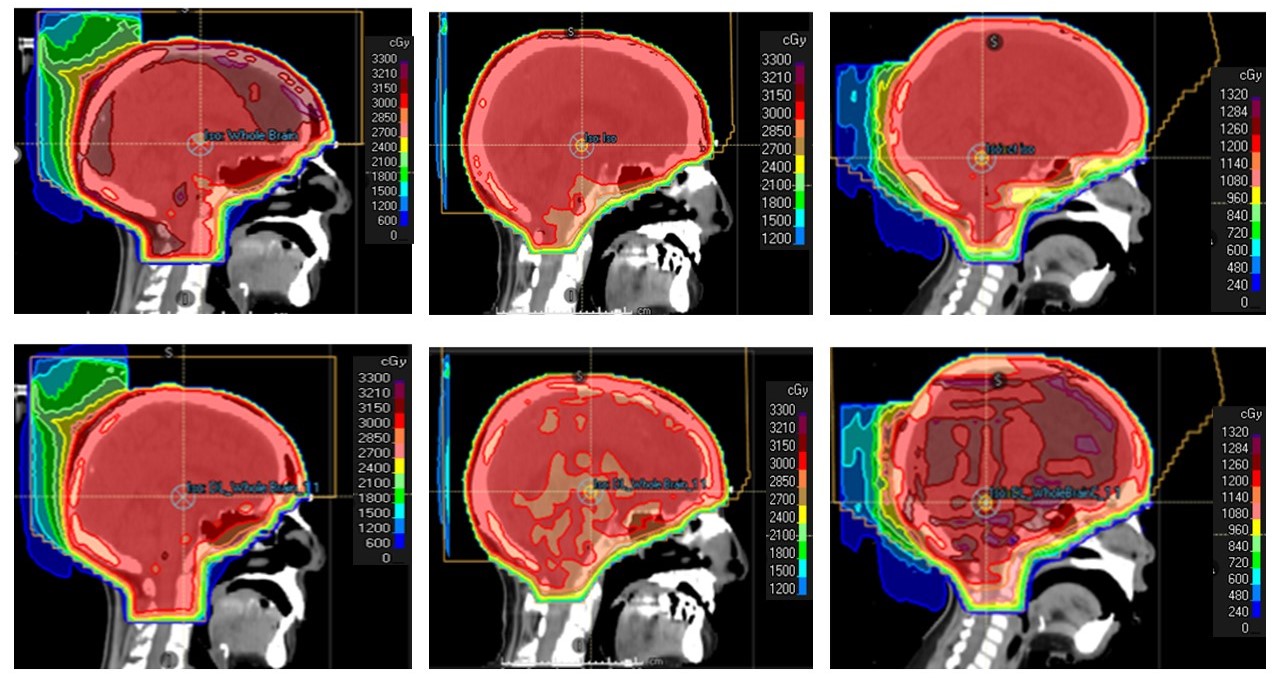}
    \caption{Examples of clinically approved plans (top) vs plans generated by the Hyperparameter Prediction module (bottom) for approved (left), approved with minor edits (middle), and rejected (right) cases. The middle red region indicates the CTV.}
    \label{fig:sample}
\end{figure}

Table~\ref{tab:dl_macro_results} presents the macro-averaged precision, recall, and F1 scores for the three DL components in the Hyperparameter Prediction module: target coverage priority, number of subfields, minimum segment MU per fraction, and minimum segment area, demonstrating consistently high performance across tasks. Figure~\ref{fig:sample} shows a comparison of three example clinical plans with three corresponding DL-optimized plans, of which one was approved immediately by the physics evaluator, one approved with minor edits, and one rejected.

\vspace{0.05cm}
\begin{table}[htbp]
\begin{center}
\caption{Macro-averaged performance (mean $\pm$ standard deviation) of the three DL modules based on five-fold cross-validation.}
\label{tab:dl_macro_results}
\begin{tabular}{p{5cm}|p{3cm}|p{3cm}|p{3cm}}
\hline
\hline
\textbf{Module} & \textbf{Precision} & \textbf{Recall} & \textbf{F1 Score} \\
\hline
Target Coverage Priority (4 classes) 
    & 0.89$\pm$0.03 & 0.87$\pm$0.04 & 0.88$\pm$0.03 \\
\hline
Number of Subfields (3 classes) 
    & 0.90$\pm$0.03 & 0.88$\pm$0.03 & 0.89$\pm$0.03 \\
\hline
MU per Fraction and Area (4 classes) 
    & 0.92$\pm$0.02 & 0.90$\pm$0.03 & 0.91$\pm$0.02 \\
\hline
\textbf{Overall Macro-average} 
    & \textbf{0.90$\pm$0.03} & \textbf{0.88$\pm$0.03} & \textbf{0.89$\pm$0.03} \\
\hline
\hline
\end{tabular}
\end{center}
\end{table}

Table~\ref{tab:dose_comparison} presents the dosimetric comparison between the plans generated by the Hyperparameter Prediction module and the clinical plans. Considering a threshold level of $p=0.05$, no statistically significant differences were observed across all evaluated metrics based on the t-test. We further present plan evaluation by the human medical physicist in Table~\ref{tab:qualitative_evaluations}. 11 of 15 (73.3\%) DL-plans were acceptable as is, 3 (20\%) required minor edits, and 1 (6.7\%) was unacceptable due to excessive hotspots and relatively high lens dose.

In terms of computation time, the inference time of the Hyperparameter Prediction module was consistently under one minute on average. The total end-to-end runtime of the Auto-FiF workflow was approximately seven minutes, which included DICOM data import from RayStation, geometric feature extraction, DL module inference, and Auto-FiF plan generation. Importantly, this entire process was fully automated and triggered with a single mouse click, requiring no manual intervention once initiated. Although Auto-FiF planning for WBRT is considered a relatively straightforward task for experienced human planners, the typical manual planning time is around 15 minutes, during which planners must iteratively adjust hyperparameters and verify plan quality. In contrast, the developed automated workflow not only reduced planning time but also streamlined the planning process by eliminating manual steps. This improvement highlights the potential of the developed system to enhance clinical efficiency while maintaining consistent plan quality.

\vspace{1cm}
\begin{table}[htbp]
\begin{center}
\caption{Comparison of dosimetric metrics between plans generated by the Hyperparameter Prediction module (DL plan) and the clinical plans.}
\label{tab:dose_comparison}
\begin{tabular}{p{3cm}|p{3cm}|p{2cm}|p{2cm}|p{2cm}}
\hline
\hline
\textbf{Structure} & \textbf{Dose Metric} & \textbf{Clinical Plan (Gy)} & \textbf{DL Plan (Gy)} & \textbf{p-value} \\  
\hline
Brain (CTV) & D1\% & 31.7±4.4 & 31.4±4.0 & 0.59 \\   
\hline
Brain (CTV) & D99\% & 29.9±5.5 & 29.7±5.3 & 0.84\\  
\hline
Eye\_L & Mean & 10.1±4.9 & 9.8±4.4 & 0.54 \\  
\hline
Eye\_R & Mean & 10.9±5.2 & 10.5±5.1 & 0.91 \\  
\hline
Lens\_L & Mean & 3.2±1.1 & 3.2±1.0 & 0.80 \\  
\hline
Lens\_R & Mean & 3.3±1.1 & 3.3±1.1 & 0.79 \\  
\hline
\hline
\end{tabular}
\end{center}
\end{table}


\begin{table}[htbp]
\begin{center}
\caption{Qualitative plan evaluations by a human medical physicist.}
\label{tab:qualitative_evaluations}
\begin{tabular}{p{0.5cm}|p{0.5cm}|p{0.5cm}|p{0.63cm}|p{2.5cm}|p{3cm}|p{3.5cm}}
\hline
\hline
\multicolumn{4}{c|}{\textbf{Plans with Scores}} & \multicolumn{3}{c}{\textbf{Clinical Acceptance}} \\  
\hline
5 & 4 & 3 & $\leq$ 2 & Acceptable & Acceptable with Minor Edits & Better than starting from scratch \\  
\hline
11 & 3 & 1 & 0 & 11 (73\%) & 3 (20\%) & 1 (7\%) \\  
\hline
\hline
\end{tabular}
\vspace{0.2cm}
\begin{minipage}{0.9\linewidth}
\footnotesize \textit{Note.} No plans received a score below 3.
\end{minipage}
\end{center}
\end{table}

\subsubsection{Explainability Analysis} 

\begin{figure}[htbp]
    \centering
    \vspace{1cm}    \includegraphics[width=\linewidth,height=0.45\textheight,keepaspectratio]{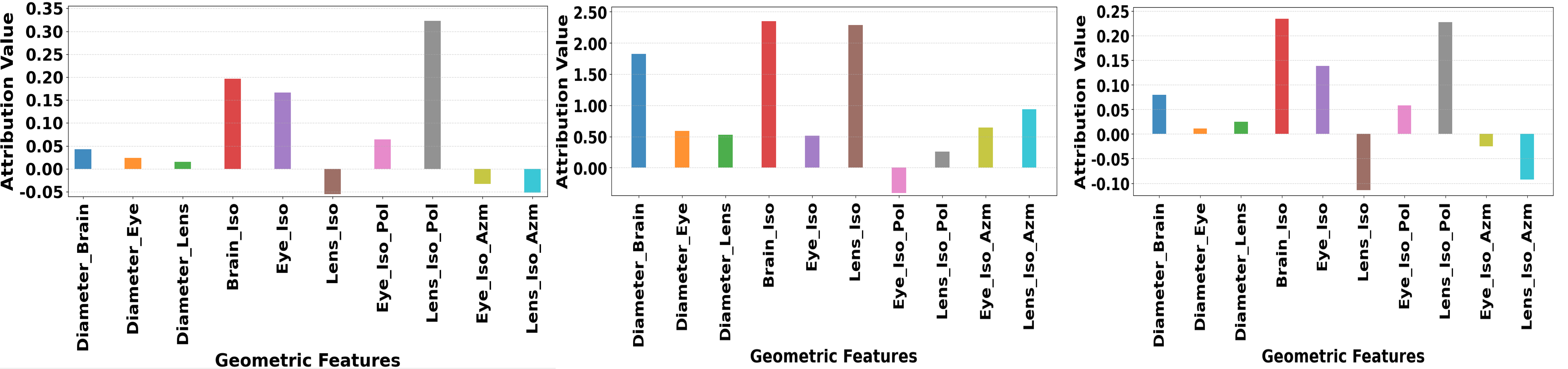}
    \caption{Visualizing Feature Contributions to Model Predictions for Target Coverage Priority (Left), Number of Subfields (Center), and Minimum Segment Settings (Right) Using Integrated Gradients Algorithm.}
    \label{fig:Importance}
\end{figure}

A strength of our model is explainability, which allowed us to gain insights about prediction behaviors and results. Figure~\ref{fig:Importance}(left) illustrates the feature contributions to the model's predictions regarding target coverage priorities. The top three contributing features were \textit{Lens\_Iso\_Pol}, \textit{Brain\_Iso}, and \textit{Eye\_Iso}. These findings suggest that target-coverage priority is influenced by how the relative positioning of the isocenter, brain, and eyes determines the balance the algorithm must achieve between ensuring sufficient target coverage and sparing the ocular structures. First, the distances from these structures to the isocenter were found to be important. Although \textit{Lens\_Iso} was not ranked among the top three features, it is strongly correlated with \textit{Eyes\_Iso}, and therefore its contribution is implicitly represented. Second, the polar angle of the eyes, \textit{Lens\_Iso\_Pol}, was identified as a significant feature. This approximately reflects the lateral spread of the eyes relative to the isocenter, thereby influencing target coverage along the beam direction.

Figure~\ref{fig:Importance}(center) shows the feature contributions to the DL model’s predictions for the number of subfields. The top three contributing features were \textit{Brain\_Iso}, \textit{Lens\_Iso}, and \textit{Diameter\_Brain}. From the treatment-planning perspective, the size of the brain is particularly influential, as it dictates the number of subfields required to achieve a sufficiently homogeneous dose coverage of the CTV. Additionally, the distances between both the brain and the lens and their respective offsets relative to the isocenter play an important geometric role in shaping dose uniformity. Similar to the previous case, \textit{Lens\_Iso} and \textit{Eyes\_Iso} are expected to exhibit strong correlation, resulting in the latter’s effect being implicitly represented through the former.


Finally, Figure~\ref{fig:Importance}(right) presents the feature contributions across all minimum segment settings (MU per fraction, segment area): (2, 4), (4, 6), (6, 9), and (8, 10). The top three contributing features were found to be \textit{Brain\_Iso}, \textit{Lens\_Iso\_Pol}, and \textit{Eye\_Iso}. These may suggest that segmentation settings depend on the spatial relationships between the isocenter, brain, and ocular structures, as closer eye or lens proximity requires finer segment shapes to control dose around these sensitive regions.



\subsection{Conversation module}

\begin{figure}[htbp]
    \centering
    \includegraphics[width=0.8\linewidth]{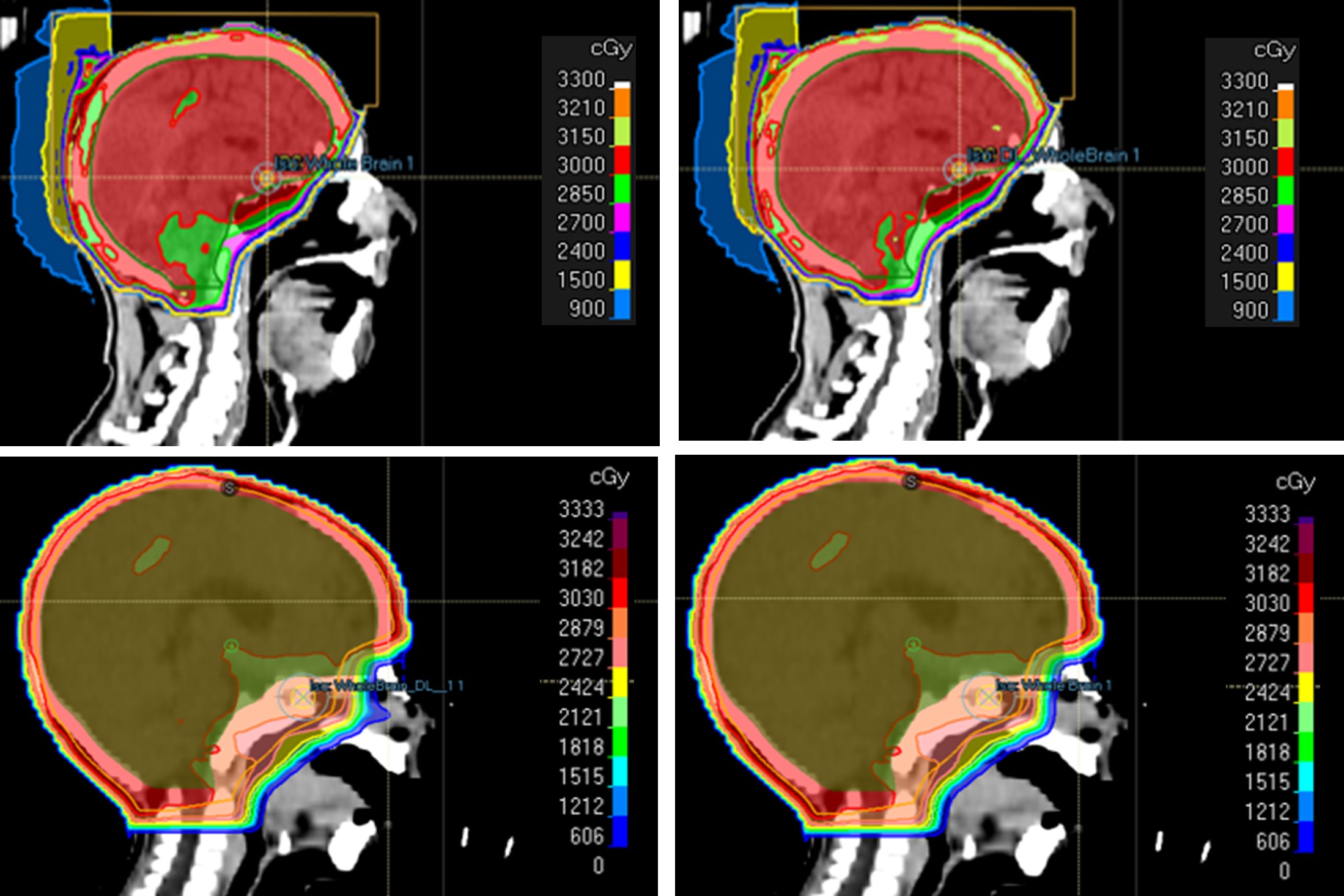}
    \caption{Comparison of the original DL-generated plans (left) with the corresponding revisions based on human feedback through the Conversation module (right). In the first case, the initial DL plan shows suboptimal target coverage, while in the second case, the initial plan delivers high doses to the left and right eyes.}
    \label{fig:conversationSample}
\end{figure}

Figure~\ref{fig:conversationSample} presents two cases where a medical physicist provided plan refinement suggestions using natural speech through the Conversation module, comparing the original DL-generated plans (left) with their revisions after feedback (right). In the first case (top-left), the initial DL plan showed suboptimal target coverage with a D95\% of 28.5 Gy, below the prescription dose of 30 Gy, with mean OAR doses of 6.8 Gy (EyeR), 5.9 Gy (EyeL), 4.2 Gy (LensR), and 4.0 Gy (LensL). The feedback was: “The target coverage doesn’t look great — I’d definitely want to improve that part. The rest seems alright.” In response to the feedback through the Conversation module, the Target Coverage Priority setting increased from 50\% to 75\%, strengthening the optimizer’s emphasis on target coverage. After revision, as shown in the top-right subfigure of Figure~\ref{fig:conversationSample}, the D95\% improved to 30.2 Gy, while the corresponding mean OAR doses were 7.0 Gy, 6.1 Gy, 4.3 Gy, and 4.1 Gy, respectively. 

In the second case (bottom-left), the DL-generated plan had adequate target coverage with D95\% = 30.9~Gy, but elevated mean doses to the EyeR (23.3~Gy) and EyeL (19.6~Gy), while the doses were 6.1~Gy for the LensR and 8.4~Gy for the LensL. The feedback provided through the Conversation module was: “The mean doses to the right and left eyes look a bit too high — I’d suggest lowering them. Target coverage seems fine.” In response, the Number of Subfields setting increased from 2 to 3, providing greater modulation flexibility to reduce the eye doses. After revision as shown in the bottom-right subfigure of Figure~\ref{fig:conversationSample}, target coverage remained stable (D95\% = 30.1~Gy), while the requested reduction in eye doses was successfully achieved—EyeR and EyeL decreased to 18.9~Gy and 15.6~Gy, respectively—and the lens doses were further lowered to 5.6~Gy (LensR) and 3.9~Gy (LensL). 

It is worth noting that these improvements cannot be achieved through simple normalization, which merely scales all doses uniformly; instead, they result from targeted re-optimization guided by the Conversation module. Together, these examples demonstrate the module’s ability to translate informal human input into actionable plan refinements.

\section{Discussions}
In this study, we proposed an end-to-end DL framework designed to automatically decide the hyperparameter settings of the Auto-FiF algorithm within the RayStation TPS. This approach enabled automation of WBRT treatment planning, reducing the need for time-consuming manual adjustments and expert intervention. The results from qualitative analysis, as performed by a qualified medical physicist, showed that the DL-optimized plans were clinically acceptable in the majority of the test cases, with only minor adjustments required to align dose distributions with clinical standards. These findings strongly supported the clinical feasibility of the proposed DL framework for WBRT, highlighting its potential to streamline and enhance the planning process. 

The Conversation module, powered by Whisper and GPT-4o, enabled the seamless translation of unstructured human feedback into structured instructions that can directly guide treatment planning. The primary purpose of this module was to address the persistent need for human intervention in the treatment planning process. Even when the initial automated plan was of high quality, it cannot guarantee that all clinical objectives are consistently met. Traditionally, resolving such issues requires iterative back-and-forth communication between physicians and dosimetrists, which can be time-consuming and prone to interpretation variability. By contrast, the Conversation module allows physicians to provide direct verbal instructions. This capability enabled real-time, natural-language interaction with the TPS, effectively bypassing the conventional physician–planner feedback loop. The module served as a critical bridge between clinical expertise and automated planning, ensuring that plans can be refined interactively and efficiently to meet individualized clinical goals, which greatly enhanced the flexibility and robustness of the automated planning workflow. This feature is expected to be even more important in scenarios that requires frequent and timely plan refinement and decision making, such as online adaptive radiotherapy\citep{LIMREINDERS2017994,li2013automatic}, as well as in settings that involve complex treatment-planning optimization models\citep{zarepisheh2014multicriteria,nguyen2022advances,cunha2020brachytherapy}.

The explainability analysis using integrated gradients further strengthened the transparency of the proposed DL-based WBRT workflow in clinical settings. In DL-based models for healthcare, it is often challenging to understand how the model arrives at specific medical decisions, which can create uncertainty and hesitation at clinical adoption\citep{jia2020clinical}. By applying integrated gradients, we were able to identify which geometric features most strongly influenced the model’s decisions in determining specific hyperparameter settings of the Auto-FiF algorithm. This transparent DL-based WBRT workflow can foster trust in system adoption and clinical integration. Moreover, the explainability analysis can help detect potential errors or biases in the model, enabling continuous refinement and ensuring safety, reliability, and effectiveness in treatment planning.

The increasing incidence of brain metastases, attributed to advancements in MRI imaging, improved screening of at-risk patients, and enhanced systemic therapies, has placed a significant strain on healthcare resources to manage this scenario \cite{Singh2022}. The proposed end-to-end DL-based workflow has the potential to streamline the WBRT workflow, allowing healthcare resources to be allocated to more critical tasks. Meanwhile, we recognize that WBRT treatment planning is a relatively simple task. We chose this task to demonstrate the overall feasibility of our framework. To our knowledge, this is the first study to integrate explainable AI into a TPS workflow together with conversational feedback, allowing physicians to refine and finalize plans seamlessly through natural language input. Although this proof-of-concept focused on WBRT, the framework is general and can be extended to more complex disease sites and treatment techniques, where treatment planning is more challenging. We expect that extending the current framework to other scenarios can bring similar advantages, such as reduction in planning time, minimizing human error, and ensuring consistency in treatment planning, further enhancing the overall efficiency and reliability of the process.



\section{Conclusion}

To streamline WBRT treatment planning, we developed an automated planning framework that integrates two key components: a Hyperparameter Prediction module and a Conversation module. The Hyperparameter Prediction module leverages supervised DL to automatically infer optimal Auto-FiF parameter settings from patient-specific geometric features, enabling efficient generation of clinically acceptable plans without manual parameter tuning. Complementing this, the Conversation module translates human feedback into structured planning objectives, thereby allowing planners to interact directly with the TPS to refine plans through natural language instructions. Tested on a cohort of 15 WBRT patients, the Hyperparameter Prediction module produced clinically acceptable plans in 93\% of cases (73\% directly and 20\% with minor edits) with no statistically significant differences ($p > 0.05$) in dosimetric metrics compared with clinical plans, while the Conversation module effectively facilitated interactive plan refinement when adjustments were required. The explainability analysis of the Hyperparameter Prediction module offered valuable insights into the model’s decision-making process. Together, these components enable a fully automated, one-click planning workflow that not only reduces planning time but also maintains flexibility for planner input and oversight.

\section*{Author Contributions}

A.J. and X.J. conceived of the presented idea. A.J., A.Q., G.A., and X.H. collected and prepared data. A.J. and A.Q. developed the the computational algorithms and implementations. X.H. and Y.G contributed to evaluation methodology and results interpretation. All authors verified the methods and results. A.J. and X.J. drafted the initial manuscripts. All authors critically reviewed and approved the final manuscript.

\section*{Conflict of Interests}
The authors declare no conflict of interests.

\section*{Acknowledgments}
This work was supported in part by NIH grants R37CA214639, R01CA227289, R01CA254377. 


\section*{References}
\addcontentsline{toc}{section}{\numberline{}References}
\vspace*{-10mm}





\bibliography{reference} 




\bibliographystyle{medphy}

\end{document}